\documentclass{pasj02} 
\usepackage[switch,mathlines]{lineno} % add line number to manuscript
\usepackage{natbib} 
\usepackage{color}

\jyear{2026}
\newcommand{\beq}{\begin{equation}}
\newcommand{\eeq}{\end{equation}}
\newcommand{\beqn}{\begin{eqnarray}}
\newcommand{\eeqn}{\end{eqnarray}}
\newcommand{\pfrac}[2]{ \biggl(\dfrac{#1}{#2}\biggr) }

\defcitealias{Okuzumi24}{Paper I}

\begin{document}

\title{Shifting the snow line in protoplanetary disks by planet-induced multiple spiral shocks}

\author{
Shizu \textsc{Shimizu}\altaffilmark{1}\altemailmark\email{shimizu.s.d77a@m.isct.ac.jp},
Ryosuke T. \textsc{Tominaga}\altaffilmark{1}\orcid{0000-0002-8596-3505},
Satoshi \textsc{Okuzumi}\altaffilmark{1}\orcid{0000-0002-1886-0880},
and
Takayuki \textsc{Muto}\altaffilmark{2}
}

\altaffiltext{1}{Department of Earth and Planetary Sciences, Institute of Science Tokyo, 2-12-1 Ookayama, Meguro, Tokyo 152-8551, Japan}
\altaffiltext{2}{Division of Liberal Arts, Kogakuin University, 1-24-2 Nishi-Shinjuku, Shinjuku, Tokyo 163-8677, Japan}

%% !!! Select 3 to 5 words from PASJ's key words !!! 
%% List of Key Words: https://academic.oup.com/pasj/pages/Pasj_Keywords 
%% "\KeyWords{ }" always has to be placed before ``\maketitle'' 


\maketitle

\begin{abstract}
The snow line location provides important constraints on rocky planet formation. Recent magnetohydrodynamical simulations suggest that Joule heating associated with gas accretion is so inefficient that the snow line can be located inside 1 au around a solar-type star. This challenges in-situ formation of the terrestrial planets in the solar system unless other internal heating processes are present.
Previous studies have proposed that spiral shocks induced by an embedded planet potentially shift the snow line outward.
In this study, we investigate the impact of shock heating induced by a giant planet on the location of the snow line in two steps. First, we perform two-dimensional hydrodynamical simulations of a disk with a giant planet to model shock heating rates. 
We construct an empirical formula for total entropy jumps due to multiple spiral shocks interior to the planetary orbit. 
Second, we calculate the radial temperature profile based on the thermal balance with the modeled shock heating rates. 
We find that shock heating can substantially increase the disk temperature in inner regions far from the planet when the planetary mass and disk optical depth are sufficiently high. For example, assuming a fixed surface density profile and an opacity of $\kappa = 1~\rm cm^2\,g^{-1}$, a planet with one-third of Jupiter's current mass at 5 au can shift the snow line outward beyond 1 au. Our results suggest that the early formation of proto-Jupiter in the solar nebula may have enabled rocky planetesimals and protoplanets to form in the present-day terrestrial planet region without invoking additional water removal processes.
\end{abstract}

%\pagewiselinenumbers 

%%%%%%%%%%%%%%%%%%%%%%%%%%%%%%%%%%%%%%%%
\section{Introduction}\label{sec:intro}
%%%%%%%%%%%%%%%%%%%%%%%%%%%%%%%%%%%%%%%%
The terrestrial planets, such as Earth, are characterized by their low water content. The mass of Earth's current ocean accounts for $0.02~\%$ of its total mass \citep{CharetteSmith10}. 
The total water content is at most $2~\%$ even when the water contained in the Earth's interior is taken into account \citep[e.g.,][]{Fei17, Tagawa21}. In contrast, comets formed in the outer solar system are thought to contain $10~\%$ water by mass \citep[e.g.,][]{Rotundi15, FUlle17, Choukroun20}. It remains a fundamental question in planet formation under what conditions such water-poor planets and water-rich small bodies form.

To clarify the origin of the low water content of the terrestrial planets, it is crucial to understand the physical conditions in protoplanetary disks and their formation process.
The disk temperature determines whether water exists as a solid or a gas. The boundary where the gas temperature is equal to the sublimation temperature of water is called the snow line. The snow line divides the disk into the inner rocky region and the outer icy region. Outside the snow line, water exists as solid ice, constituting up to $50\%$ of the dust mass \citep{Lodders2003}. Therefore, rocky planets have conventionally been considered to form inside the snow line \citep[e.g.,][]{Hayashi81}.
Recent studies have proposed that rocky bodies can form outside the snow line. For instance, \citet{Lichtenberg2019} suggested that planetesimals can be dehydrated by $\rm ^{26}Al$ heating. However, because of the short half-life of $\rm ^{26}Al$, planetesimals may still remain water-rich depending on the timing of their formation. Furthermore, even if planetesimals are born dry, they can evolve into water-rich planets by capturing icy pebbles \citep{Sato16}. \citet{Johansen2021} proposed that water ice in accreted pebbles sublimates in the hot atmosphere of protoplanets. In this case, the water content of the planet is determined by the amount of ice accreted before the temperature of the planetary envelope reaches the sublimation temperature. 
However, a recycling flow is required to efficiently return water vapor to the disk; otherwise, the planet acquires a large amount of water. Therefore, it remains important to explore a simple scenario in which rocky planets form inside the snow line.

From a theoretical perspective, the disk temperature distribution, and hence the location of the snow line, remains highly uncertain because it depends on the available heating sources and the disk's optical thickness. The classical solar nebula model of \citet{Hayashi81} assumed an optically thin disk and placed the snow line at 3 au. However, the assumption of low optical thickness is inadequate, at least during the early stages of planet formation, when dust grains keep the disk optically thick. In such a disk, the interior temperature can be either higher or lower than in an optically thin disk, depending on the presence or absence of strong internal heating sources. If the disk is heated solely by stellar irradiation, the interior temperature can be about a factor of two lower than in an optically thin disk \citep[e.g.,][]{Kusaka70,Chiang97}, placing the snow line inside 1 au \citep[e.g.,][]{SL00}. 
Accretion heating at the disk midplane would push the snow line out to about 1 au over the first million years of disk evolution \citep[e.g.,][]{Davis05,Garaud07,Oka11,Bitsch15}. However, this scenario has been challenged by recent magnetohydrodynamical (MHD) simulations, which demonstrate that Joule heating associated with accretion is concentrated near the disk surface and therefore contributes little to heating the disk interior \citep{Mori19,Mori25}. In such disks, the snow line migrates inside 1 au within the first million years of disk evolution \citep{Mori21,Kondo23}. This emerging picture implies that water ice would have been abundant at Earth's current orbit in the solar nebula even during the early stages of planet formation.

In this study, we focus on shock heating induced by a giant planet as a new heating source. A planet embedded in a disk excites spiral density waves through gravitational interaction with the disk \citep{GoldreichTremaine79}. As the density waves propagate, they develop into shock waves and heat the disk \citep{Rafikov16}. \citet{Ziampras20} investigated the thermal structure of a planet-hosting disk using 2D hydrodynamical simulations including viscous heating, stellar irradiation, and radiative cooling. They showed that a giant planet raises the temperature in its vicinity and inside its orbital radius.
In the solar nebula, the isotopic dichotomy suggests that Jupiter formed at an early stage of disk evolution \citep{Kruijer20, Kleine20}. 
Therefore, shock heating induced by Jupiter could have significantly influenced the water content of the solar system bodies.

Toward a systematic understanding of shock heating, \citet{Ono2025} quantified the shock heating rate by measuring the jump in specific entropy along streamlines (entropy jump) using 2D hydrodynamical simulations of disks with an embedded planet. However, their simulations only cover the region beyond half of the planet's orbital radius. Assuming that the giant planet is located at Jupiter's current orbital radius (5 au), their computational domain is outside the terrestrial planet region of the solar system. Simulations covering further inner regions are necessary to understand how Jupiter affects the temperature of the terrestrial planet-forming region. Some density waves induced by a giant planet become shock waves far from the planet \citep{BaeZhu2018b}. These shock waves potentially contribute to the disk heating in the inner region.

In this paper, we aim to reveal how efficiently the shock waves induced by a giant planet heat the inner disk regions and change the location of the snow line. 
To achieve this, we conduct 2D hydrodynamical simulations that extend domains further inward than in \citet{Ono2025} and calculate the shock heating rates. We apply the obtained shock heating rates to wind-driven accretion disks where accretion heating is inefficient. Then, we evaluate the shift of the snow line location in the disk with shock heating.

The outline of this paper is as follows. In section \ref{sec:simulation}, we present the numerical simulations of a disk with a giant planet to model the entropy jumps due to multiple shock waves induced by the planet. In section \ref{sec:temperature}, we show how shock heating changes disk temperature profiles and the location of the snow line. We also discuss implications for rocky planet formation. In section \ref{sec:conclusions}, we summarize our conclusions and future work.

\section{Hydrodynamical simulations}\label{sec:simulation}
In this section, we present 2D hydrodynamical simulations to evaluate shock heating rates in the inner region of protoplanetary disks. In subsection \ref{subsec:simu_method}, we describe the numerical method. We present the results of our simulations in subsection \ref{sec:simu_kekka} and provide an empirical formula for the radial profile of the shock heating rates in subsection \ref{subsec:fitting}. In subsection \ref{subsec:simu_discus}, we discuss the validity of the empirical formula and the simulations.

\subsection{Method} \label{subsec:simu_method}
We carry out 2D hydrodynamical simulations for protoplanetary disks in polar coordinates $\left( r,\phi\right)$. We adopt a frame co-rotating with a planet around a central star. Our measurements of shock heating rates utilize the streamline analysis in \citet{Ono2025}. We refer readers to \citet{Ono2025} for the details of the numerical methods.

\subsubsection{Basic equations}

The equations of continuity, motion, and energy are given by 
\begin{align}
     \frac{d\Sigma}{d t}  &= -\Sigma\nabla \cdot \boldsymbol{v}, \\
     \Sigma\frac{d  \boldsymbol{v}}{d t} &= - \nabla p + \nabla\cdot\Pi +\boldsymbol{g} + \Sigma\left(\Omega^2_\mathrm{p}\boldsymbol{r}-2\Omega_\mathrm{p}\boldsymbol{n}_z\times\boldsymbol{v}\right), \\
    \frac{d \Sigma e}{d t} &= -\gamma\Sigma e\nabla  \cdot \boldsymbol{v} + \frac{1}{2\nu\Sigma}\left(\Pi^2_{rr}+2\Pi^2_{r\phi}+\Pi^2_{\phi\phi} \right) \notag   \\&~
    + \frac{2\nu\Sigma}{9}\left(\nabla\cdot \boldsymbol{v}\right)^2 + Q_{\mathrm{relax}},
\end{align}
where $\Sigma$ is the gas surface density, $\boldsymbol{v}$ is the gas velocity, $p$ is the vertically integrated pressure, $\boldsymbol{g}$ is the gravitational acceleration, $\Omega_\mathrm{p}$ is the angular velocity at the planetary orbit $r_\mathrm{p}$, $\Pi_{ij}$ is the $(i,j)$ component of the viscous stress tensor, $\nu$ is the kinematic viscosity. We employ the $\alpha$-viscosity model \citep{Shakura73} and adopt $\nu=\alpha c_{\rm s}^2\Omega_{\rm K}^{-1}$, where $\alpha$ is the viscosity parameter, $c_{\rm s}$ is the sound speed, and $\Omega_{\rm K}$ is the Keplerian angular velocity. The viscous stress tensor and the gravitational acceleration are given by equations (3) and (6) of \citet{Ono2025}, respectively, where $M_*$ is the stellar mass and $M_\mathrm{p}$ is the planetary mass. We adopt a smoothing length $\epsilon$ to avoid the divergence of the planet's gravitational potential. We set $\epsilon=0.05~r_{\rm p}$, which is comparable to the disk scale height\footnote{There is an error in the description of the smoothing length in \citet{Ono2025}. The actual smoothing length adopted in their simulations was not $\epsilon = 0.6r_\mathrm{H}$ but $\epsilon=0.05r_{\rm p}$. 
}.
We introduce the thermal relaxation term $Q_{\mathrm{relax}}$ to model radiative cooling \citep{Gammie01}:
\begin{align}
    Q_{\mathrm{relax}}&= -\frac{\Sigma\left(e-e_\mathrm{i}\right)\Omega_\mathrm{i}}{\beta},
\end{align}
where $e=p/[\Sigma(\gamma-1)]$ is the internal energy per unit mass, and $\Omega_\mathrm{i}$ is the initial angular velocity \citep[see equation (10) of][]{Ono2025}. The adiabatic index $\gamma$ is fixed at 1.4 in this study.
The relaxation term forces $e$ to relax to the initial internal energy $e_\mathrm{i}$ on a timescale of $\beta / \Omega_\mathrm{i}$.
We assume a spatially constant $\beta$ and set $\beta=100$ throughout the disk.

Following \citet{Ono2025} and \cite{Okuzumi26}, we neglect viscous heating associated with Keplerian shear flow to isolate shock heating. This technical choice is motivated by the recent finding that accretion heating has a minor contribution to the midplane heating \citep{Mori19,Mori25}.

\subsubsection{Numerics}
We use the Athena++ code \citep{stone20} for our simulations.
We adopt the second-order Runge--Kutta method for time integration, the second-order piecewise linear method for spatial reconstruction, and the Harten--Lax--van Leer--contact approximate Riemann solver \citep{Toro1994} to calculate the gas flux. We also employ the orbital advection scheme \citep{Masset2000}. The computational domain has a smaller inner boundary than in \citet{Ono2025}, covering $[r_{\mathrm{in}},r_{\mathrm{out}}] = [0.1r_{\mathrm{p}}, 3.9r_{\mathrm{p}}]$ in the radial direction and $[-\pi,\pi]$ in the azimuthal direction. We employ 1024 cells in the radial direction and 1548 cells in the azimuthal direction. The computational grid is uniformly spaced in the azimuthal direction and logarithmically spaced in the radial direction to ensure that the aspect ratio of each cell is approximately unity. In this setup, the grid size at planetary orbit is $\approx 3.6\times 10^{-3}~r_\mathrm{p}$. As the radial boundary conditions, we fix the physical variables to the initial values. We apply the periodic boundary condition in the azimuthal direction. Following \citet{Ono2025} and \citet{Okuzumi26}, we set wave-damping zones \citep{valborro06} to reduce wave reflection at the radial boundaries.

The initial disk surface density and the sound speed are given by 
\begin{align}
    \Sigma _ \mathrm{i} \left( r \right) &= \Sigma _0 \pfrac{r}{r_\mathrm{p}}^{\xi}, \\
    c_{\rm i} \left( r \right) &= r_{\rm p}\Omega_{\rm p}h_{\rm p} \pfrac{r}{r_\mathrm{p}}^{-\xi/2-3/4},\label{eq:ci}
\end{align}
where $\Sigma_0$ and $h_{\rm p}$ are the initial surface density and aspect ratio at the planetary orbital radius, and $\xi$ is the power-law slope of the initial surface density. The radial dependence of $c_{\rm i}$ is chosen so that mass accretion rates, $\dot{M} = 3\pi \nu \Sigma$, are radially constant. 
The initial gas velocity $\boldsymbol{v}_\mathrm{i}=\left( v_{r\mathrm{i}}, ~v_{ \phi \mathrm{i} } \right)$ is given by equations (9) and (10) of \citet{Ono2025}. The initial radial velocity is consistent with the steady accretion model \citep{Lyndenbell74}. The initial azimuthal velocity ensures radial force balance. 
The planet mass gradually increases from zero to $M_\mathrm{p}$ over the first 50 orbits \citep[see equation (15) of][]{Ono2025}.

We present seven runs as listed in table \ref{table:parameter} to construct an empirical formula for the shock heating rate. In these runs, we adopt $\beta=100$ throughout the disk.
For $\beta>10$, the damping of spiral arms due to thermal relaxation is negligible \citep{MirandaRafikov2020a,MirandaRafikov2020b}. We fix the value of $\beta$ because the shock heating rate is independent of $\beta$ in such highly adiabatic disks \citep{Ono2025}. 
We investigate the dependence of the shock heating rate on the planet mass $M_\mathrm{p}=qM_*$, the viscosity parameter $\alpha$, the initial aspect ratio at the planetary orbital radius $h_\mathrm{p}$, and the radial dependence of the surface density $\xi$. The simulations are carried out for 4000 orbits. RUN 1 is our fiducial model. 

\begin{table}[t]
    \begin{center}
    \begin{tabular}{c c c c c c}
        \hline
        Run name & $q$ & $\alpha$ & $h_{\rm p}$ & $\xi$ & $M_\mathrm{p}/ M_\mathrm{{th}}$ \\ 
        \hline
        RUN 1 & $10^{-3}$   & $10^{-3}$   & 0.05 & $-1$   & 8.0 \\
        RUN 2 & $10^{-3.5}$ & $10^{-3}$   & 0.05 & $-1$   & 2.5 \\
        RUN 3 & $10^{-4}$   & $10^{-3}$   & 0.05 & $-1$   & 0.8 \\
        RUN 4 & $10^{-3}$   & $10^{-3.5}$ & 0.05 & $-1$   & 8.0 \\
        RUN 5 & $10^{-3}$   & $10^{-3}$   & 0.07 & $-1$   & 2.9 \\
        RUN 6 & $10^{-3}$   & $10^{-3}$   & 0.05 & $-0.5$ & 8.0 \\
        RUN 7 & $10^{-3}$   & $10^{-3}$   & 0.05 & $-1.5$ & 8.0 \\
        \hline
    \end{tabular}
    \end{center}
    \caption{List of parameter sets}
    \label{table:parameter}
\end{table}

\subsubsection{Measuring the entropy jumps across spiral shocks}\label{sec:measurement}
We calculate the shock heating rate using the jump in entropy across shock waves. The shock heating rate per unit area is given by \citep[see][]{Rafikov16, Ziampras20, Ono2025} 
\begin{align}
     Q_{\mathrm{sh}} = \frac{\Delta \left(\Sigma e\right)}{\Delta t} = \frac{p_1}{\Delta t \left( \gamma-1\right)}\delta,
\label{eq:Qsh}
\end{align}
where $\Delta t=2\pi/|\Omega_\mathrm{K}(r)-\Omega_\mathrm{p}|$ is the time interval between shock crossings.
Here, $\delta$ is the dimensionless entropy jump given by
\begin{equation}
     \delta \equiv  \mathrm{exp}\left(\Delta S/c_{\rm v}\right) -1
    = \frac{p_2/\Sigma^{\gamma}_2}{p_1/\Sigma^{\gamma}_1} -1,
    \label{eq:delta}
\end{equation}
where the subscripts 1 and 2 represent the pre-shock and post-shock values, respectively, $\Delta S$ is the specific entropy jump across the shock, and $c_{\rm v}$ is the specific heat.

We calculate dimensionless entropy jumps along streamlines. When the system reaches quasi-steady state at the end of the simulations, the streamlines coincide with the trajectories of gas fluid elements. 
This streamline analysis enables us to measure the entropy increase due to the shock because the entropy is conserved along a streamline in the adiabatic limit.
We draw streamlines using the velocity field at the final simulation time. The dimensionless entropy jump is determined based on the minimum and maximum entropy values along each streamline. {The detection limit of $\delta$ in the present study is $\simeq1\times10^{-4}$, which is set by apparent entropy changes irrelevant to spiral shocks.}

A single planet can excite multiple spirals \citep[e.g.,][]{Kley99,DongZhuRafikov15,Zhu15,FungDong15,BaeZhu2018a,BaeZhu2018b}, which steepen into shocks at different distances from the planet. 
{If multiple shocks exist along a single streamline, we can identify them as spatially separated entropy jumps. For further analysis, we measure entropy jumps and their locations for each shock. In most cases in the present simulations, we can evaluate total entropy jumps as the sum of the largest and second-largest entropy jumps. We find that neglecting smaller jumps changes total values by only a few tens percent or less. We also neglect small entropy jumps less than 10\% of the largest $\delta$ for each streamline to avoid overestimation of $\delta$ due to apparent entropy changes.}
We draw streamlines at 100 logarithmically spaced radii to derive radial profiles of the dimensionless entropy jumps. We exclude the damping zones and the gap region from the measurement.

\subsection{Results}\label{sec:simu_kekka}

\subsubsection{Fiducial case}
\begin{figure*}
\begin{centering}
    \includegraphics[width=0.537\hsize]{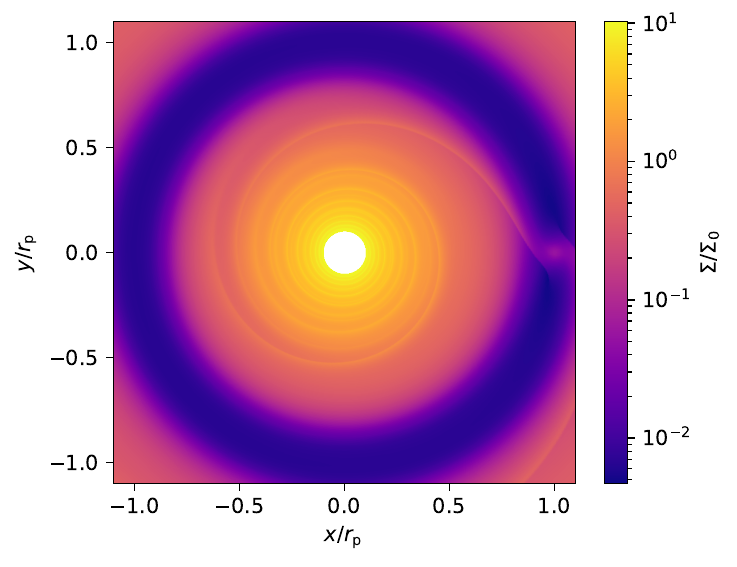}
    \includegraphics[width=0.46\hsize]{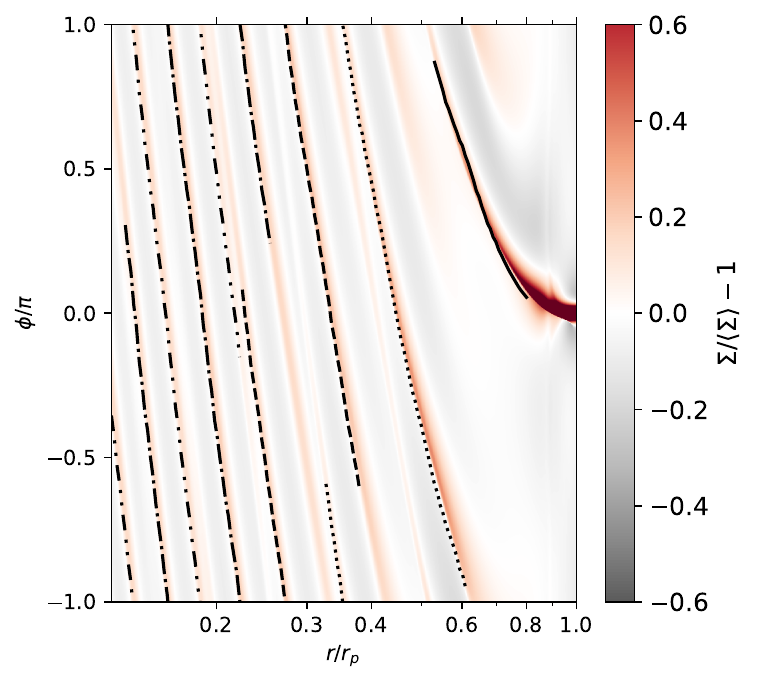}
\end{centering}
    \caption{Left panel: Snapshot of the disk surface density distribution from RUN 1. The planet is located at $(x,y)=(r_{\rm p},0)$. Multiple spiral arms are present inside the planetary orbit ($r=r_{\rm p}$). The low surface density in the planet's vicinity represents the gap formed by the planet. The central white region is out of the computational domain. Right panel: Relative deviation of the surface density from its azimuthal average $\left<\Sigma\right>$, plotted in the $r$--$\phi$ plane to highlight the inner spiral arms. The solid, dotted, dashed, dash-dotted, and dash-dot-dotted lines mark the shocks, identified by their entropy jumps, associated with the primary, secondary, tertiary, quaternary, and quinary arms, respectively.}
    \label{fig:sigma}
\end{figure*}

We begin by presenting the results of the fiducial run (RUN 1). We confirm that the disk has nearly reached a quasi-steady state by the end of the simulation, as the surface density profile varies by less than 25\% during the final 1000 orbits. 
The left panel of figure \ref{fig:sigma} shows the 2D distribution of the gas surface density. The planet carves a deep gap in its vicinity. The planet also excites multiple spiral density waves. 
Inside the planetary orbit, we detect five spirals that emerge at different positions. We call them the primary, secondary, tertiary, quaternary, and quinary arms in order of increasing distance from the planet's orbit. These arms are more clearly visible in the right panel of figure~\ref{fig:sigma}, which shows the surface density perturbations in the inner disk plotted in the $r$--$\phi$ plane.
\citet{BaeZhu2018b} showed that a planet can excite up to five spiral arms in the inner disk. However, they detect only three spiral arms with the same parameter set as in RUN 1. This is likely because their simulations are limited to the range of $r>0.2 r_{\rm p}$ while our computational domain extends down to $r=0.1 r_\mathrm{p}$. 

\begin{figure}[t]
\begin{center}
\includegraphics[width=\linewidth]{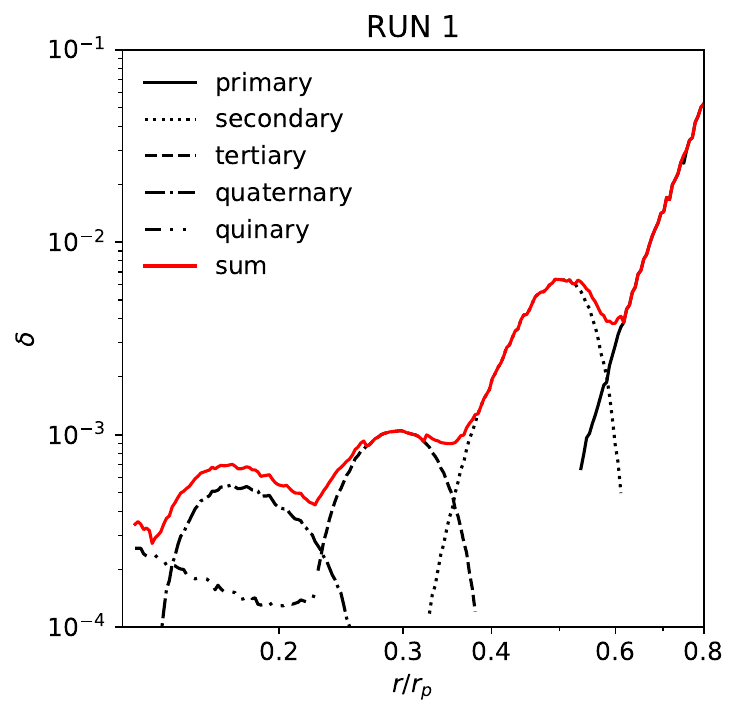}
\end{center}
\caption{Dimensionless entropy jumps $\delta$ (equation~\eqref{eq:delta}) of multiple spirals interior to the planetary orbit, plotted as a function of the distance from the central star, $r$, for RUN 1. The black lines represent the contributions from the primary to quinary spiral shocks, whereas the red line shows their sum.
}
\label{fig:delta}
\end{figure}

Figure \ref{fig:delta} shows the radial profiles of the dimensionless entropy jumps, defined by equation~\eqref{eq:delta}, for the five inner spiral arms. 
The primary arm is already shocked in the immediate vicinity of the planet, whereas the secondary, tertiary, quaternary, and quinary arms develop into shocks at $r/r_{\rm p} \approx$ 0.6, 0.4, 0.3, and 0.2, respectively. 
As shown in figure \ref{fig:delta}, the primary shock dominates the total entropy jump around the gap, whereas the secondary and higher-order shocks dominate in the disk interior to the gap. Importantly, although the higher-order shocks are weaker than the primary shock, the cumulative entropy jump across the tertiary, quaternary, and quinary shocks {decreases from $r \approx 0.4 r_{\rm p}$ to $r \approx 0.1 r_{\rm p}$ more gradually than that across the primary and secondary shocks}.

\subsubsection{Parameter dependence}\label{sec:parameter}
We now investigate the parameter dependence of the entropy jump profiles. The left panel of figure \ref{fig:delta_allrun} shows radial profiles of the entropy jump from all runs. 

RUNs 2 and 3 adopt planetary masses that are three and ten times lower than that in RUN 1, respectively. Because low-mass planets induce weak shock waves, the entropy jumps in these runs are smaller than in RUN 1. 
In RUN 2, the primary, secondary, and tertiary waves clearly form. We also identify the quaternary wave although it is less clear than in RUN 1. {In contrast, the entropy jump profile from RUN 3 clearly shows only the primary and secondary shocks. The higher-order shocks in RUN 3 are weak, and their entropy jumps do not create peaks in the radial profile of total entropy jumps, although they can be identified as relative deviations of the surface density from its azimuthal average as in figure \ref{fig:sigma} (see appendix \ref{appendix:lowmass}).  
The measured $\delta$ for the higher-order shocks at $r\lesssim0.3r_{\rm p}$ is comparable to the detection limit (subsection \ref{sec:measurement}), which is due to apparent entropy changes concealing shock-induced entropy jumps.
Thus, the total entropy jumps might be less accurate at $r\lesssim0.3r_{\rm p}$ in RUN 3 and overestimated.}

RUN 4 employs an $\alpha$ value three times lower than in RUN 1. 
We find that the $\delta$ values of the primary, secondary, and tertiary shocks are smaller than those in RUN 1.
For the primary shock, our result is consistent with those of 
\citet{Ono2025} and \citet{Okuzumi26}, who showed that the primary shock's $\delta$ profile
scales as $\sqrt{\alpha}$. Our results further confirm that {the similar positive correlation} also applies to the secondary and tertiary shocks {(see also the next subsection for quantitative comparison}). 
In contrast, the $\delta$ values of the quaternary and quinary waves in RUN 4 are comparable to those in RUN 1. This could be because the entropy jumps for these distant shocks have not yet fully converged within 4000 orbits. In fact, the viscous relaxation timescale in RUN 4 is longer than in RUN 1 owing to the lower $\alpha$ value {(see appendix \ref{appendix:long_sims}).}

RUNs 5, 6, and 7 illustrate the dependence of the $\delta$ profile on the sound speed profile (see equation~\eqref{eq:ci}). In RUNs 5 and 6, the sound speed in the inner region is higher than in RUN 1, causing the secondary and higher-order waves to steepen into shocks at greater distances from the planet. The opposite trend is seen in RUN 7, where the sound speed in the inner region is lower than in RUN 1. 
{This is because the wave propagation speed through the disk is determined by the local sound speed \citep[e.g., see][]{Rafikov02}, and thus the locations of the secondary and higher-order shocks are also scaled with the sound speed.}\footnote{{Another possible reason is the difference in relative strength of planetary gravity and gas pressure, which is often measured by $q/h_{\rm p}^3$ in the context of planet-disk interaction \citep[e.g.,][and see also equation (\ref{eq:M_th})]{LinPapaloizou1993,Dong11}. We consider that this is minor in the parameter space explored in the present study. This is because the locations of the peaks in RUNs 1 and 2 are similar despite the difference in $q/h_{\rm p}^3$.}}
These trends are consistent with the $h_{\rm p}$-dependence of the locations of multiple gaps shown in previous studies \citep{BaeZhu2018b,Dong2018,Zhang2018}. This similarity is reasonable and stems from the fact that the entropy jumps and gaps are linked by the shock formation. At the peak of the entropy jump, gas accretion is further promoted by {transfer of angular momentum} through the shock. This results in a local decrease in the surface density, thereby opening a gap at the peak of $\delta$ under certain conditions.

\subsection{Empirical formula for the total entropy jump profile}\label{subsec:fitting}
Based on the simulation results presented in the previous subsection,  we now provide an empirical expression for the entropy jump profile as a function of the input parameters of the simulations, ($q, \alpha, h_{\rm p}$). 

\begin{figure*}
    \begin{centering}
    \includegraphics[width=0.5\hsize]{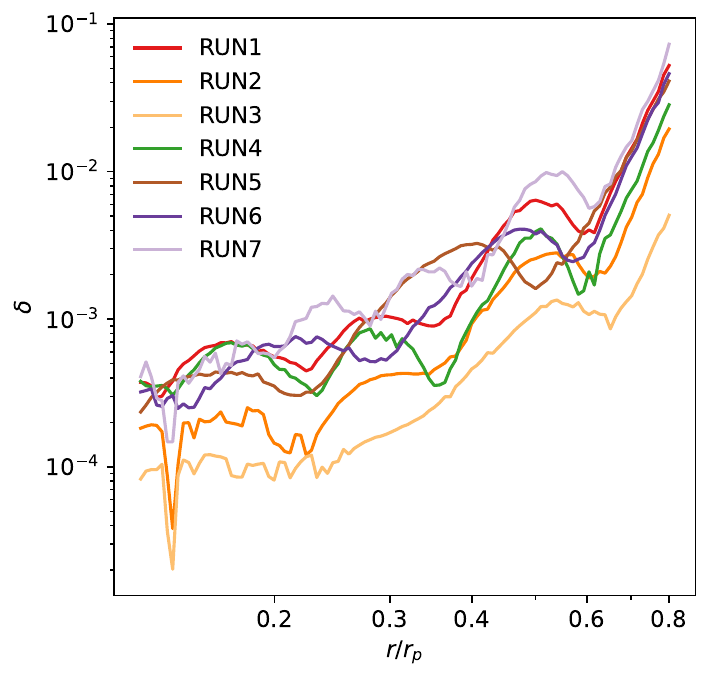}
    \includegraphics[width=0.5\hsize]{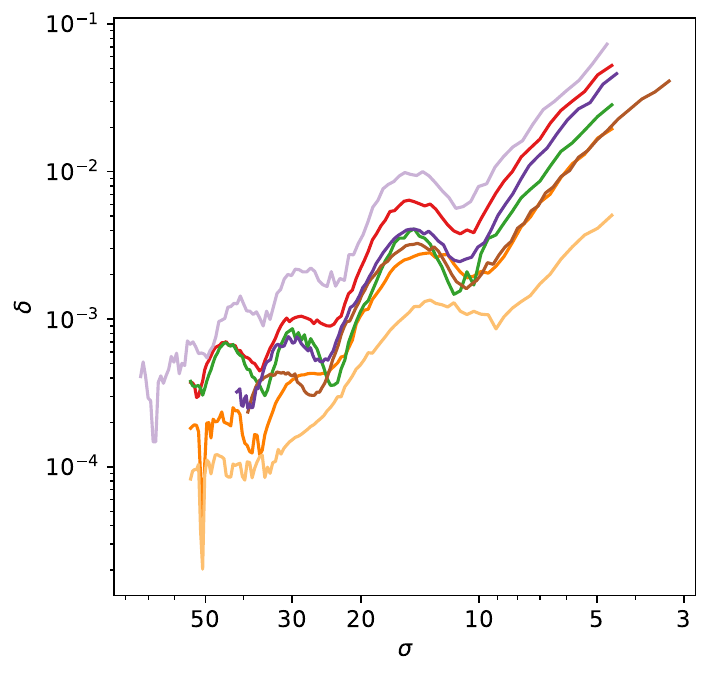}
    \end{centering}
    \caption{Left panel: Total dimensionless entropy jump as a function of the distance from the central star, $r$, from all runs. Right panel: Same as the left panel, but showing the entropy jump profiles as a function of the scaled distance from the planetary orbit, $\sigma$ (equation (\ref{eq:sigma})).
    }
    \label{fig:delta_allrun}
\end{figure*}

We begin by introducing the scaled distance from the planet.
As discussed in the previous subsection, the distances between the planet's orbit and the shocks scale with the sound speed.
This trend motivates us to introduce the scaled distance $\sigma$ defined as
\begin{align}
    \sigma = \int_r^{r_\mathrm{p}} \frac{dr}{H(r)} \label{eq:sigma}, 
\end{align}
where $H(r) = c_{\rm s}/\Omega$ is the scale height. 
The right panel of figure \ref{fig:delta_allrun} shows the entropy jump profiles as a function of $\sigma$ instead of $r$. Here, we use the initial sound speed distribution and Keplerian angular velocity to determine $\sigma$ (i.e., $H(r)= c_\mathrm{i}/\Omega_{\rm K}$). In the simulations presented here, the temperature remains close to its initial value, and the scale height changes by no more than 20\%.
One can see that the peaks of the entropy jumps associated with the secondary,  tertiary, and quaternary shocks are consistently found at $\sigma \simeq 15$, $30$, and 45, respectively, whenever they are detected.

\citet{Ono2025} and \citet{Okuzumi26} provided empirical formulas for $\delta$ associated with the primary shock. 
However, higher-order shocks can also collectively contribute to secular heating of the inner disk. It is therefore desirable to derive a formula for the {\it total} entropy jump.  \citet{Ono2025} and \citet{Okuzumi26} showed that the scaling $\delta \propto \sqrt{\alpha /h_{\rm p}}M_{\rm p}/M_{\rm th}$ applies to the primary shock, where 
\begin{align}
    M_{\rm th} \equiv h_{\rm p}^3 M_* \label{eq:M_th}
\end{align}
is the thermal mass \citep{LinPapaloizou1993,GoodmanRafikov01}.
Assuming that the same scaling applies to higher-order spiral shocks, we propose the following  empirical formula for the total entropy jump as a function of $\sigma$:
\begin{align}
    \delta = \mathcal{A} \sqrt{\frac{\alpha}{h_{\rm p}}}\frac{M_{\rm p}}{M_{\rm th}}\sigma^\mathcal{B}(\mathcal{C}\mathrm{exp}\left( -\mathcal{D}\sigma)~\mathrm{sin}(\sigma / \mathcal{E})+1\right),
    \label{eq:fitting_entjump}
\end{align}
where $(\mathcal{A,~B,~C,~D,~E})=(10^{-0.4},-1.7,~0.7,~0.02,~2.2)$ are  fitting parameters.
In this new formula, each peak of the sinusoidal function corresponds to the peak of the $\delta$ profile for a different spiral arm.

\begin{figure*}
    \begin{centering}
    \includegraphics[width=0.5\hsize]{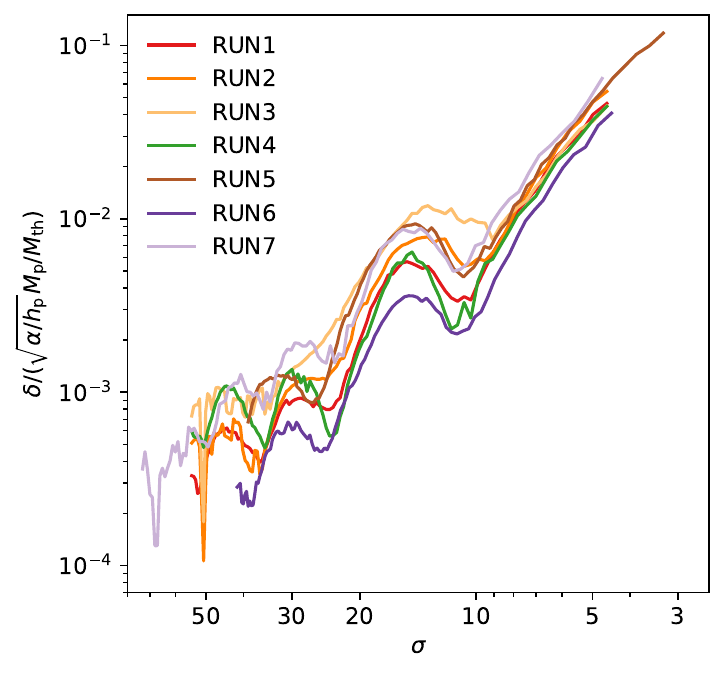}
    \includegraphics[width=0.5\hsize]{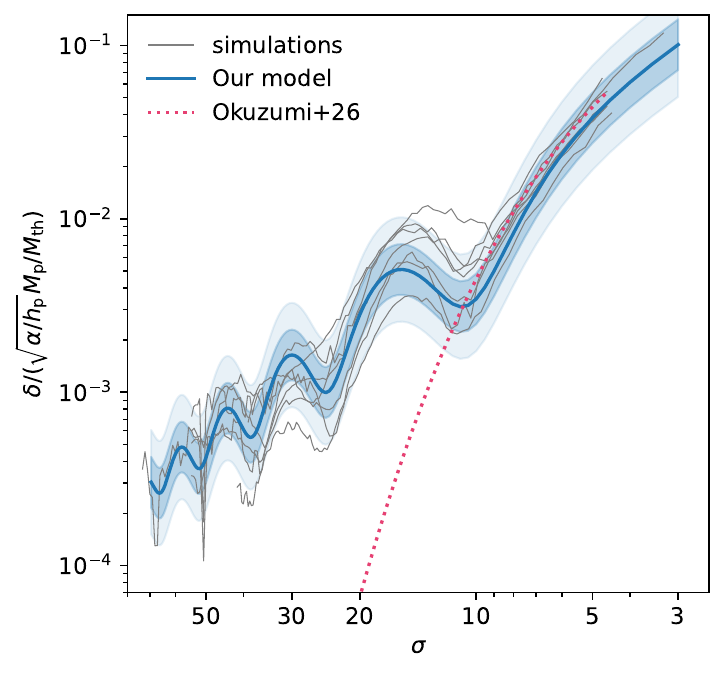}
    \end{centering}
    \caption{Left panel: Same as the right panel of figure~\ref{fig:delta_allrun}, but with the total entropy jumps scaled by $\sqrt{\alpha/h_{\rm p}}M_{\rm p}/M_\mathrm{th}$. 
    Right panel: Same as the left panel, but with the simulation results (thin solid lines) overlaid by the empirical formula (thick solid line; equation~\eqref{eq:fitting_entjump}). The dark and light shaded bands indicate deviations by factors of 1.4 and 2 from the empirical formula, respectively. The dotted line represents the empirical formula for the entropy jump of the inner primary shock from \citet{Okuzumi26}.
    }
    \label{fig:delta_scaling}
\end{figure*}

Figure \ref{fig:delta_scaling} compares the radial profiles of the scaled total entropy jump,  $\delta/(\sqrt{\alpha /h_{\rm p}}M_{\rm p}/M_{\rm th})$, from all our simulations with the prediction of equation~\eqref{eq:fitting_entjump}.  As shown in the right panel of figure \ref{fig:delta_scaling}, our empirical formula reproduces the total entropy jump to within a factor of two over $5 \lesssim \sigma \lesssim 70$. For comparison, we also plot the entropy jump formula for the inner primary shock derived by \citet{Okuzumi26}. As expected, the previous formula underpredicts the measured total entropy jump at $\sigma \gtrsim 10$, where the secondary and higher-order spiral shocks dominate.

\subsection{Validity of the empirical formula} \label{subsec:simu_discus}
The empirical expression for the dimensionless entropy jump derived in the previous subsection (equation~\eqref{eq:fitting_entjump}) enables us to estimate the rate of disk heating by planet-induced spiral shocks, $Q_{\rm sh}$ (equation~\eqref{eq:Qsh}), interior to the planetary orbit over the parameter space explored in this study (table~\ref{table:parameter}). Here, we discuss the validity and limitations of our entropy jump modeling. 

\begin{figure}
    \begin{centering}
    \includegraphics[width=\linewidth]{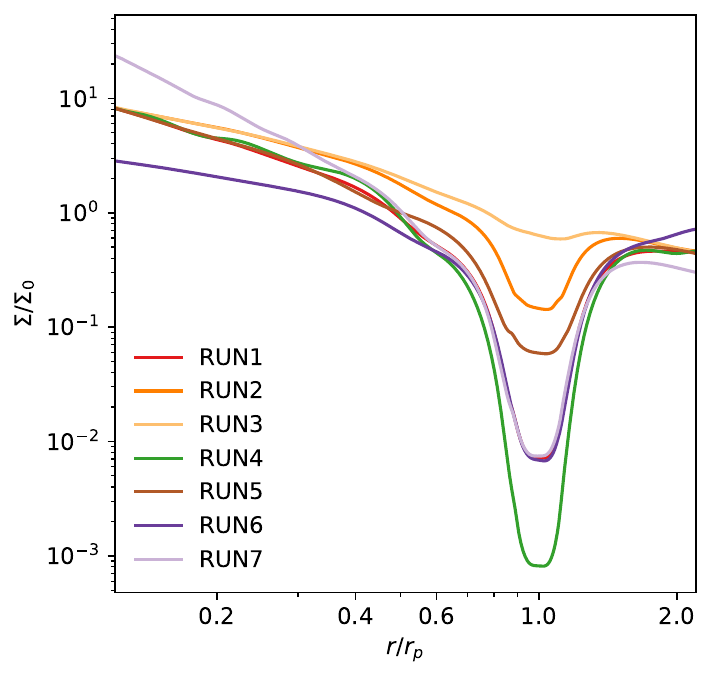}
    \end{centering}
    \caption{Comparison of the radial surface density profiles from all runs.
    }
    \label{fig:sigma_allrun}
\end{figure}

As shown in the left panel of figure \ref{fig:delta_scaling}, the scaled values of the entropy jump differ by a factor of two among the runs. 
We first discuss the origin of this deviation. The scaled entropy jumps associated with the secondary waves in RUNs 2, 3, 5, and 7 are a factor of two higher than those in the other RUNs.
The secondary waves start to shock at $r/r_{\rm p}\simeq0.6$ in RUN 1 and within $r/r_{\rm p}\simeq 0.5-0.7$ in the other RUNs (figures \ref{fig:sigma} and \ref{fig:delta_allrun}).
The surface densities at the inner edge of the gap are higher in RUNs 2, 3, and 5 than in the other RUNs (figure \ref{fig:sigma_allrun}), while those at the secondary shock locations remain within the same order of magnitude. We thus speculate that the deviation in the scaled entropy jumps is partly due to the difference in the surface density at the gap edge rather than that at the shock locations. This seems reasonable from a viewpoint of wave excitation. The density waves excited by the planet transport angular momentum, and then shock waves deposit angular momentum onto the disk. This angular momentum transport may depend on the surface density at the inner gap edge. 
This may bring the dependence of $\delta$ on the surface density at the inner gap edge because the entropy jump is linked to the angular momentum deposition \citep{Rafikov16}. 

However, in RUN 7, the surface density at the inner gap edge is almost the same as that in RUN 1. 
Therefore, the surface density may not be the unique factor leading to the deviation in the entropy jumps. 
Another possible reason is the effect of the local aspect ratio $h$ at the shock locations. We discuss the impact of this factor-of-two uncertainty on the disk temperature in section \ref{sec:temperature}.

{
Our empirical formula might be less accurate for higher-order shocks induced by a low-mass planet (e.g., $r\lesssim 0.3r_{\rm p}$ in RUN 3). This is because of the detection limit of our streamline analysis. According to RUNs 1, 2, and 3, the detection limit becomes a concern for $\delta$ of tertiary and higher-order shocks when $q$ is less than approximately $1\times10^{-4}$ for $h_{\rm p}=0.05$. The radial profile of total entropy jumps across these high-order shocks appears to be monotonic rather than sinusoidal, in contrast to the functional form of our formula. Further studies on these limitations are warranted in our future study.
}

{
As noted in the previous section and appendix \ref{appendix:long_sims}, our empirical formula may overestimate entropy jumps in low-viscosity cases ($\alpha\sim10^{-3.5}$, RUN 4) because a longer-term run shows lower $\delta$. This is particularly the case for entropy jumps due to tertiary and higher-order shocks. Practically, in low-viscosity cases, one would have to take into account wind-driven accretion to model protoplanetary disk evolution of typical mass accretion rates \citep[for recent reviews, see][]{Lesur23,Bai26}, which will be addressed in our forthcoming paper (Tominaga et al. in prep.).
}

Our simulations are restricted to 2D. In 3D disks, shock waves are weaker than in 2D because gas expands in the vertical direction \citep{Bate2003,Lyra16}. Thus, we may overestimate the entropy jump in our 2D simulations. To obtain a more realistic estimate of the entropy jump, we need to extend our models by considering three-dimensional gas flows.

%%%%%%%%%%%%%%%%%%%%%%%%%%%%%%%%%%%%%%%%%%%%%%%
\section{Application to snow line modeling} \label{sec:temperature}
%%%%%%%%%%%%%%%%%%%%%%%%%%%%%%%%%%%%%%%%%%%%%%%
In this section, we utilize the empirical equation for the planet-induced shock heating rate to investigate the impact of shock heating on the location of the snow line in a protoplanetary disk. In subsection \ref{subsec:tsh_method}, we present the method for calculating radial temperature profiles based on a one-dimensional disk model. In subsections \ref{subsec:tsh_kekka} and \ref{subsec:implications}, we describe the results and discuss the implications for rocky planet formation, respectively.

\subsection{Method} \label{subsec:tsh_method}
To investigate the impact of shock heating on the disk temperature, we perform 1D disk temperature calculations. We consider an optically thick, wind-driven accretion disk in which accretion heating is concentrated near the disk surface and has a negligible effect on the interior temperature \citep{Mori19,Mori21,Mori25,Kondo23}. The only significant sources of interior heating are therefore stellar irradiation and planet-induced shock heating. As we are particularly interested in the impact of proto-Jupiter on the temperature structure of the solar nebula, we assume a star with solar mass and solar luminosity and a giant planet with an orbital radius of $r_{\rm p} = 5$ au.

The disk surface density profile is given by 
\begin{equation}
    \Sigma \left( r \right)= 1700 \left( \frac{r}{1 ~\mathrm{au}} \right)^{-1} \mathrm{g\,cm}^{-2},
\end{equation}
where the value at 1 au is taken from the minimum-mass solar nebula model of \citet{Hayashi81}. 
The isothermal sound speed $c_\mathrm{s}$ is given by
\begin{equation}
    c_\mathrm{s} = \sqrt{\frac{k_\mathrm{B}T_\mathrm{mid}}{m}}
    \label{eq:cs},
\end{equation}
where $k_\mathrm{B}$ is the Boltzmann constant, $T_{\mathrm{mid}}$ is the midplane temperature, and $m =2.35~m_\mathrm{H}$ is the mean molecular mass with $m_\mathrm{H}$ being the hydrogen atomic mass. 

The midplane temperature $T_{\rm mid}$ is determined by the balance between irradiation heating, shock heating, and radiative cooling. This thermal balance can be expressed as \citep{Ziampras20,Wu21} 
\begin{align}
    \left( 1+\frac{3}{4}\tau_{\mathrm{mid}}\right)^{-1}\left( F_*\mathrm{sin} \theta - 2 \sigma_{\mathrm{SB}} T^4_{\mathrm{mid}}\right)  +Q_{\rm sh} =0, 
    \label{eq:temperature}
\end{align}
where $\tau_{\mathrm{mid}}$ is the vertical optical depth of the disk midplane, $F_*$ is the energy flux of the central star's radiation, $\theta$ is the angle between the starlight and the disk's irradiation surface, and $\sigma_{\mathrm{SB}}$ is the Stefan--Boltzmann constant. In equation~\eqref{eq:temperature}, the first two terms account for irradiation heating and radiative cooling, whereas the last term represents shock heating (see equation~\eqref{eq:Qsh}).
The optical depth and irradiation flux are given by
\begin{align}
    \tau_{\mathrm{mid}} &= \frac{1}{2}\kappa\Sigma, \\
    F_* &= \frac{L_*}{4\pi r^2} ,
\end{align}
where $\kappa$ is the disk's opacity and $L_*$ is the stellar luminosity (assumed to be solar), and we have assumed that the opacity is vertically constant. 
Focusing on the inner $\sim 1~\rm au$ disk region, we fix the starlight's grazing angle to be $\sin\theta = 0.05$ throughout the disk \citep{Chiang97}.
{We note that adopting a temporally constant grazing angle excludes irradiation instability \citep[e.g.,][]{Watanabe08,Kutra24,Sudarshan26} from the present analysis.
This simplification would not change our results qualitatively because irradiation instability is stabilized when disk internal heating dominates irradiation heating \citep[e.g., see][]{Watanabe08}, which is the case of our particular focus in this study.}
The opacity $\kappa$ depends on the dust size distribution, temperature, composition, and morphology. For example, for porous dust with a size distribution up to a maximum size of $1~\rm \mu m$, the opacity $\kappa$ is several $\rm cm^2\,g^{-1}$ at a few hundred $\rm K$ \citep[e.g.,][]{semenov03}. For simplicity, we adopt a constant value of either $\kappa=0.1$ or $~1 ~\rm cm^2\,g^{-1}$.

The shock heating rate $Q_{\rm sh}$ is calculated using equation~\eqref{eq:Qsh} with the empirical expression for $\delta$, equation~\eqref{eq:fitting_entjump}. 
The pre-shock pressure $p_1$ in $Q_{\rm sh}$ is approximated by the azimuthally averaged 2D pressure, $c_{\rm s}^2\Sigma$. 
The $\alpha$ parameter in equation~\eqref{eq:fitting_entjump} is set to  $10^{-3}$, corresponding to the laminar Maxwell accretion stress found in non-ideal MHD simulations of protoplanetary disks (see, e.g., figure 10 of \citealt{Bai17}). We note that this non-zero stress does not contradict our neglect of accretion heating at the disk midplane. In weakly ionized disks, Joule dissipation tends to occur at higher altitudes than the Maxwell stress \citep[see also][]{Hirose11}.

Since $Q_{\rm sh} \propto p_1 \propto T_{\rm mid}$, we solve equation (\ref{eq:temperature}) for $T_{\rm mid}$ numerically using an iterative method. 
As our model for $Q_{\rm sh}$ is not applicable to the close vicinity of the planetary orbit, where spiral shocks are absent, our calculation only covers the region of $r \leq r_{\rm p} - H_{\rm p}$, where $H_\mathrm{p}=h_\mathrm{p}r_\mathrm{p}$ stands for a typical shocking length \citep{GoodmanRafikov01, Rafikov02}. 
We fix the value of $\sigma$ (\ref{eq:sigma}) at $r=r_{\rm p}-H_{\rm p}$ to unity and calculate those at inner radii with updated temperature. In the following, we simply adopt $h_\mathrm{p} = 0.05$ in equation \eqref{eq:fitting_entjump} and calculate the temperature profile at $r < 4.75$ au.
We note that the adopted value of $h_{\rm p}$ is higher than that determined by the balance between irradiation heating and radiative cooling (i.e., $Q_{\rm sh}=0$ in equation (\ref{eq:temperature})). The aspect ratio $h_{\rm p}$ should be lower than we assume if the gap region is optically thin, and the temperature is set by irradiation heating (e.g., see figures 6 and 7 of \citet{Ziampras20}). Adopting higher $h_{\rm p}$ gives conservative estimates for shock heating rates for a given planet mass $M_{\rm p}$ because of $\delta \propto h_\mathrm{p}^{-3.5}$ (see equations \eqref{eq:M_th} and \eqref{eq:fitting_entjump}).
We also calculate the possible uncertainty range of $T_{\rm mid}$ by assuming that  equation~\eqref{eq:fitting_entjump} has a factor-of-two uncertainty (see subsection~\ref{subsec:fitting} and the right panel of figure~\ref{fig:delta_scaling}).  
The snow line is taken to be the location at $T_{\rm mid}$ = 170 K \citep{Hayashi81}.

\subsection{Results}\label{subsec:tsh_kekka}
\begin{figure}[t]
\begin{center}
\includegraphics[width=\linewidth]{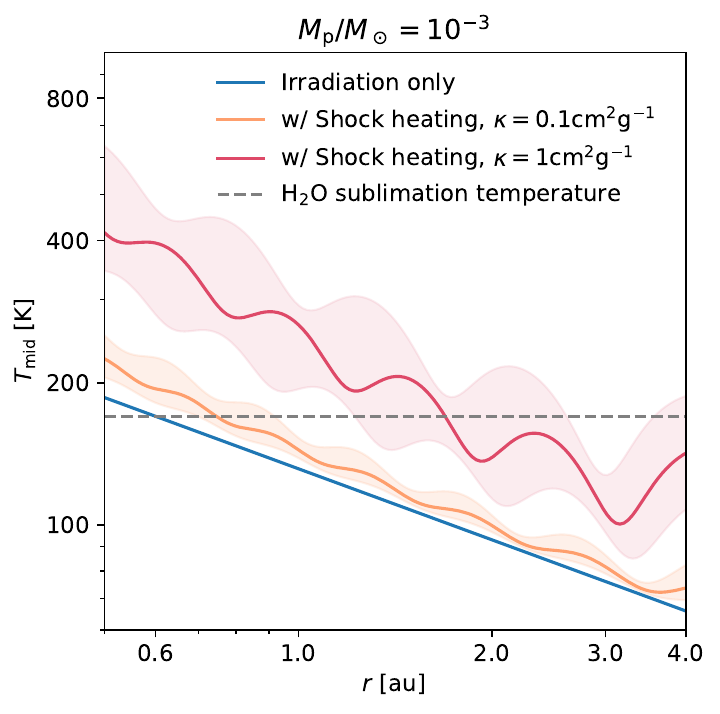}
\end{center}
\caption{Radial profiles of midplane temperature $T_{\rm mid}$ for the disk model presented in section~\ref{sec:temperature}. The bottom solid curve corresponds to the case without shock heating, whereas the upper two solid curves include heating by spiral shocks produced by a $M_{\rm p} = 10^{-3}$ (i.e., approximately Jupiter-mass) planet at $r_{\rm p} =$ 5 au, with disk opacities of $\kappa=1$ and $0.1~\rm cm^2\,g^{-1}$. The light-shaded regions around the upper two curves indicate the uncertainty in $T_{\rm mid}$ propagated from the factor-of-two uncertainty in the empirical formula for $\delta$, equation~\eqref{eq:fitting_entjump} (see subsection~\ref{subsec:fitting} and the right panel of figure~\ref{fig:delta_scaling}). The dashed line marks the sublimation temperature of water ice, assumed to be $170~\rm K$.
}
\label{fig:temp}
\end{figure}

As highlighted in section~\ref{sec:intro}, 
{a disk that is optically thick to stellar irradiation} with no internal heating source can have a snow line interior to the 1 au orbit. This is demonstrated in figure~\ref{fig:temp}, where we plot $T_{\rm mid}$ of our optically thick disk model with $Q_{\rm sh} = 0$. In this model, the snow line (where $T_{\rm mid}$ = 170 K) lies at 0.6 au. 

We now examine whether planet-induced shock heating can significantly raise the temperature of such a passively irradiated disk. We first present the results for $M_{\rm p}=10^{-3}M_\odot$ (approximately one Jupiter mass) in figure~\ref{fig:temp}. For $\kappa=1~\rm cm^2\,g^{-1}$, planet-induced shocks substantially raise the midplane temperature over a wide region interior to the planet's orbit (located at 5 au). For instance, the temperature at 1 au reaches $\sim 300$ K, about a factor of two higher than in the disk without shock heating. Strikingly, the temperature increase due to planet-induced shock heating becomes more pronounced toward smaller radii, even though $\delta$ decreases with increasing distance from the planet. This is because the midplane optical depth increases toward smaller radii, making radiative cooling progressively less efficient. A similar trend is seen in the hydrodynamical simulations of \citet[][see their figure 6]{Ziampras20}. The factor-of-two uncertainty in our empirical formula for $\delta$ translates into an uncertainty of approximately 50\% in $T_{\rm mid}$ (see figure~\ref{fig:temp}). {The uncertainty curves are phase-shifted relative to the analytical curve. This phase shift stems from the difference in sound speed. A higher value of $\delta$ leads to a higher temperature, and thus a higher sound speed. The increased sound speed makes the scaled distance from the planet $\sigma$ smaller (see equation \ref{eq:sigma}).  As a result, the peaks of $\delta$ and temperature move farther from the planet.}

The temperature increase due to shock heating shifts the snow line outward. Figure~\ref{fig:temp} shows that, once a Jupiter-mass planet forms at 5 au in a disk with $\kappa = 1~\rm cm^2\,g^{-1}$, the snow line moves to $r \approx 1.2$--2.6 au.

We note that the results presented above are highly dependent on the assumed opacity. For $\kappa=0.1~\rm cm^2\,g^{-1}$, the temperature at 1 au increases by only 20 K, and the snow line remains at $r \approx 0.7$--0.9 au even in the presence of a Jupiter-mass planet. In this case, disk heating is dominated by stellar irradiation, and the uncertainty in the empirical model of $\delta$ has only a minor effect on $T_{\rm mid}$.

\begin{figure}
    \begin{center}
    \includegraphics[width=\linewidth]{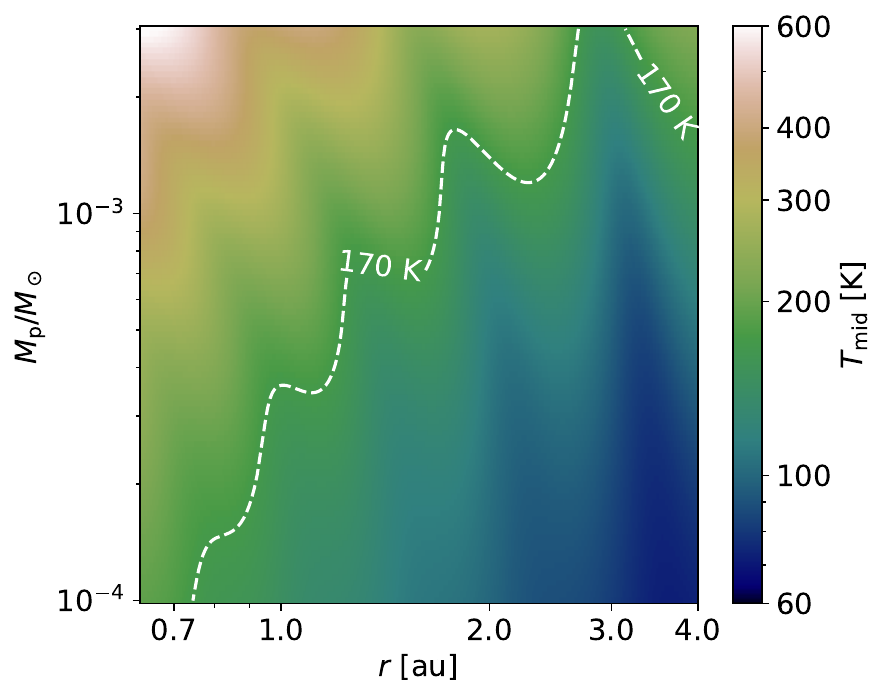}
    \end{center}
    \caption{Two-dimensional color map showing the midplane temperature $T_{\rm mid}$ as a function of orbital radius and planet mass for the disk model presented in section~\ref{sec:temperature} with $\kappa = 1~\rm cm^2\,g^{-1}$. The planet is assumed to lie at $r_{\rm p} = 5~\rm au$. The total entropy jump in the shock heating rate is evaluated using equation (\ref{eq:fitting_entjump}). The dashed line traces the water snow line ($T_{\rm mid}=170~\rm K$).
    }

    \label{fig:snowline_evolution}
\end{figure}

We also investigate the location of the snow line for different planetary masses. Figure \ref{fig:snowline_evolution} shows that the snow line moves outward as the planetary mass increases for $\kappa=1~\rm cm^2\,g^{-1}$. This is because an increase in planetary mass results in higher $\delta$ and hence higher $Q_\mathrm{sh}$. For certain narrow ranges of $M_{\rm p}$, the disk possesses multiple snow lines owing to the non-monotonic temperature distribution caused by multiple spiral shocks.

\subsection{Implications for terrestrial planet formation in the solar nebula}\label{subsec:implications}

We now put the results presented in the previous subsection in the context of rocky planet formation.
Rocky planets have conventionally been thought to form inside the snow line  \citep{Hayashi81}, while recent studies proposed some scenarios in which rocky planets form outside the snow line \citep{Lichtenberg2019, Johansen2021}. 
In the previous subsection, we showed that, when small dust grains are abundant and the disk is sufficiently optically thick, the snow line shifts outward after a giant planet forms and induces spiral shocks that heat the disk. This outward shift of the snow line extends the dry inner region. This implies that, in the presence of a giant planet, rocky planets can form over a wide disk region even under the conventional scenario that rocky planets form inside the snow line.

As shown in figure \ref{fig:snowline_evolution}, the location of the snow line varies with planet mass.
The snow line reaches 1 au when a giant planet at 5 au attains a mass of $\approx 3\times10^{-4}M_\odot$ for a disk with an opacity of $1~\rm cm^2\,g^{-1}$ and a surface density at 1 au of $1.7\times10^3 \rm ~g\,cm^{-2}$. 
{
We note that this critical planet mass is above the threshold below which the measurements of entropy jumps become less accurate because of the detection limit (see also sections \ref{sec:parameter} and \ref{subsec:simu_discus}).
This suggests that the snow line in the solar nebula could have moved beyond Earth's current orbit once proto-Jupiter grew beyond one-third of Jupiter's current mass if the disk was sufficiently optically thick}.

The efficiency of shock heating depends on the optical depth of the disk midplane. If the disk is optically thin {in the vertical direction}, shock heating becomes inefficient. The opacity is determined by the size and abundance of the remaining dust grains \citep{Pollack1994, semenov03}. These dust properties are determined by the dust growth, settling, radial drift, and planetesimal formation processes. After the formation of proto-Jupiter, the disk may already be optically thin because of dust evolution and planetesimal formation. Therefore, the timing of proto-Jupiter formation could be crucial to whether shock heating can sufficiently raise the disk temperature.

The optical depth may also vary depending on changes in disk structure due to shock heating.
As shown in figure \ref{fig:temp}, the disk temperature is locally higher at the $\delta$'s peak radius because of the higher shock heating rate. It can increase the pressure locally and modify the radial pressure profile. Resulting wavy pressure profiles may slow down the dust radial drift, causing a traffic jam \citep[e.g.,][]{Whipple72}. This dust accumulation makes the disk midplane optically thicker, and could further raise the disk temperature.
{\citet{Kato25} proposed this thermally driven dust accumulation by adopting viscous accretion heating as an internal heating source. We expect that thermally driven dust accumulation occurs even in wind-driven accretion disks, where viscous accretion heating is inefficient \citep[e.g.,][]{Hirose11,Mori19,Mori21}, if giant planets are embedded and induce shocks. This is because shock heating induced by a giant planet can be an alternative internal heating source.}
Multiple gap formation due to shock waves can also modify the pressure profile.
Angular momentum deposition to the disk gas through the shock is related to the shock heating rate \citep{Rafikov16, ArzamasskiyRafikov18, Okuzumi26}.
The gas radial velocity induced by the shocks is faster for higher $\delta$. Therefore, shock-induced accretion reduces the surface density at the peaks of the entropy jump. Whether the actual pressure at the radius where $\delta$ peaks increases or decreases remains uncertain and should also depend on the opacity. In the present model, we fixed the opacity and surface density profile for simplicity.
In future work, we will address simultaneous modeling of the temperature and surface density profiles with opacity changes.

%%%%%%%%%%%%%%%%%%%%%%%%%%%%%%%%%%%%%%%%%%%%%%%%%%%
\section{Conclusions and outlooks} \label{sec:conclusions}
%%%%%%%%%%%%%%%%%%%%%%%%%%%%%%%%%%%%%%%%%%%%%%%%%%%
We investigated the shift of the snow line due to shock heating induced by a giant planet. First, we carried out 2D hydrodynamical simulations that cover the inner regions of the protoplanetary disks to quantify the dimensionless entropy jump. Using the obtained empirical formula for the entropy jump, we calculated the radial temperature profile of a disk with a giant planet. Our key findings are summarized as follows.
\begin{enumerate}
\item Extending the computational domain inward enabled the detection of up to five shock waves. The radii where each shock wave forms are well reproduced by using the distance from the planet normalized by the local scale height, $\sigma$ (equation \eqref{eq:sigma}). The total entropy jump profiles follow a universal scaling that depends on the thermal mass $M_{\rm th}$, the dimensionless disk viscosity $\alpha$, the initial aspect ratio $h_\mathrm{p}$ at the planetary orbit, and the scaled distance $\sigma$ (equation \eqref{eq:fitting_entjump}). This empirical formula reproduces the entropy jumps measured in our simulations to within a factor of 2. {The formula may overpredict entropy jumps induced by higher-order shocks in the case of low planet masses or low viscosity (e.g., $q\lesssim 1\times10^{-4}$ or $\alpha \lesssim3\times10^{-4}$ for $h_{\rm p}=0.05$) because of the detection limit of the present streamline analysis.}

\item Planet-induced shock waves can substantially increase the disk temperature if the planetary mass and disk optical depth are sufficiently high.
For a Jupiter-mass planet at 5 au embedded in a disk with $\kappa=1~\rm cm^2\,g^{-1}$ and a surface density at 1 au of $1.7\times10^3 \rm~g~cm^{-2}$, the temperature at 1 au is approximately 150 K higher than that of the irradiation-dominated case (figure \ref{fig:temp}). 
The snow line moves outward as the planetary mass increases. When the planet grows to $\approx 0.3$ Jupiter masses, the snow line reaches 1 au (figure \ref{fig:snowline_evolution}). In the lower-opacity case of $\kappa=0.1~\rm cm^2\,g^{-1}$, shock heating hardly contributes to the disk temperature, and disk heating is dominated by stellar irradiation.
\end{enumerate}
{Our findings suggest that giant planets can significantly alter the distribution of water ice in protoplanetary disks if they form in inner regions of the disk (e.g., $\simeq5\;\rm au$) and the vertical optical depth is high enough}. In the solar nebula, early-formed Jupiter may have shifted the snow line beyond 1 au, thereby allowing rocky planetesimals and protoplanets to form in the present-day terrestrial planet region even under the conventional picture in which rocky bodies form interior to the snow line.

In future work, we will measure the dimensionless entropy jump for a wider parameter space with improvement of hydrodynamical models. 
In particular, it remains uncertain to what extent we can apply the present viscous model calculations to simulate shock heating in wind-driven accretion disks. Recent MHD simulations found that a deeper gap forms in a wind-driven accretion disk than in a viscous accretion disk with measured stress adopted \citep[e.g.,][]{Aoyama23}. As discussed in subsection \ref{subsec:simu_discus}, differences in the surface density around the primary gap might quantitatively change the entropy jumps. We will address this point in a forthcoming paper (Tominaga et al. in prep.).

While our present work focused on the disk temperature structure, our future modeling will include the evolution of dust distribution, opacity, and gas density structures. The temperature calculations in section \ref{sec:temperature} imply a potential interplay between shock heating and dust accumulation around planet-induced multiple gaps (subsection \ref{subsec:implications}). Self-consistent modeling of all these effects will shed light on the formation and compositional evolution of rocky planets, including our own.

\section*{Funding}
This work was supported by JSPS KAKENHI (grant numbers JP23K25923 and JP24KJ1041) and NINS Astrobiology Center program research (grant number AB0705).

%%%%%%%%%%%%%
\begin{ack}
We thank Hidekazu Tanaka and Shota Sato for fruitful discussions, and Tomohiro Ono for providing us with the simulation code, and an anonymous reviewer for constructive comments.
\end{ack}
%%%%%%%%%%%%%

\bibliographystyle{apj}
\bibliography{References}

\appendix 
\section{Higher-order shocks for low-mass planets} \label{appendix:lowmass}
{Detection of spiral shocks based on entropy jumps is more difficult for lower planet masses. As mentioned in section \ref{sec:parameter}, this is the case in RUN 3, and higher-order shocks are so weak that they do not form peaks in the radial profile of total entropy jumps. Such weak shocks are more easily identified in the density profile than in the entropy profile because entropy jumps is proportional to the cube of density jumps \citep[e.g.,][]{Landau1959}. Here, we present density variations to show the existence of higher-order shocks in RUN 3.}

\begin{figure}
    \begin{center}
    \includegraphics[width=\linewidth]{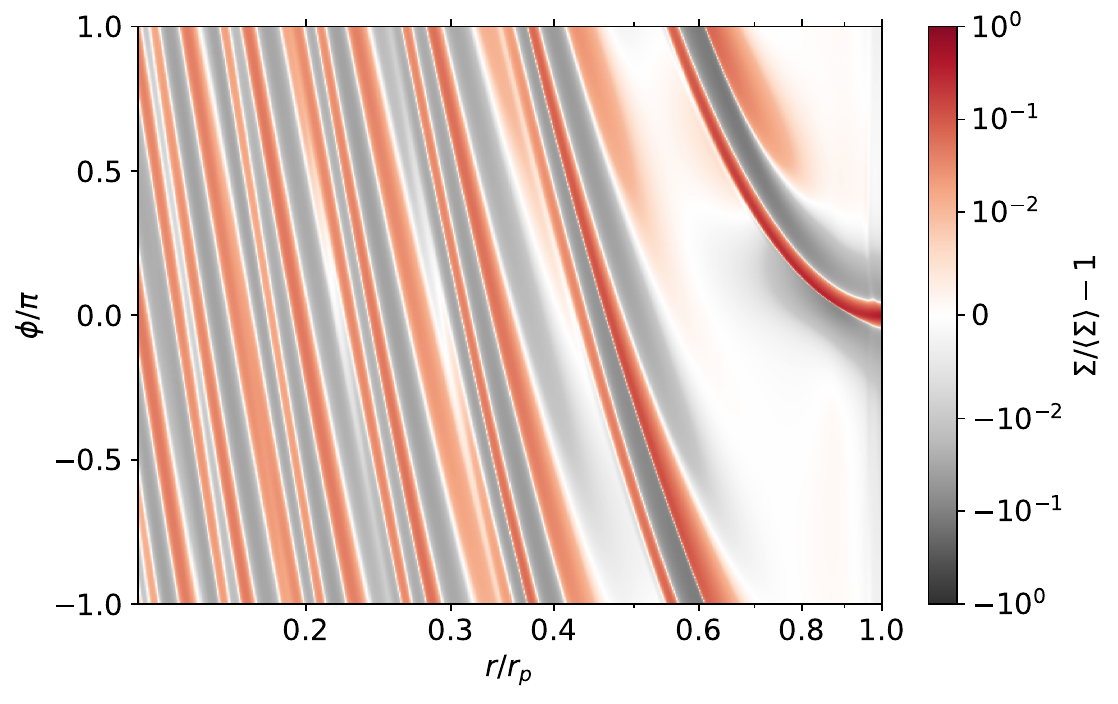}
    \end{center}
    \caption{{Same as the right panel of figure \ref{fig:sigma}, but from RUN 3. We use a logarithmic color bar to highlight small perturbations due to higher-order spirals.}
    }
    \label{fig:dsigma_q4}
\end{figure}

{
Figure \ref{fig:dsigma_q4} shows relative deviations of the surface density from its azimuthal average from RUN 3. The quaternary and quinary shock start developing approximately at $(r/r_{\rm p},\phi/\pi)\simeq(0.4,0.2)$ and $(0.2, 0.5)$, respectively. The quinary shock is very close to the tertiary shock, which is launched from $r/r_{\rm p}\simeq0.4-0.5$. This is similar to RUN 1 (see the right panel of figure \ref{fig:sigma}).
}

\section{Convergence of entropy jumps in low-viscosity case} \label{appendix:long_sims}

{
Here, we present a long-term simulation of RUN 4 to discuss convergence of entropy jumps. As noted in section \ref{sec:parameter}, the entropy jump due to tertiary, quaternary, and quinary shocks in RUN 4 has not yet converged within 4000 orbits, in contrast to RUN 1. 
The viscous relaxation timescale depends on viscosity $t_{\rm visc}\propto\alpha^{-1}$. In RUN 4, the viscous parameter $\alpha$ is three times lower than in RUN 1. 
Assuming that the planet-induced structures evolve on a timescale proportional to the viscous relaxation timescale, we can expect a quasi-steady state in approximately 12000 orbits for RUN 4.
Therefore, we extend the simulation time to 16000 orbits.}

\begin{figure}[t]
    \begin{center}
    \includegraphics[width=\linewidth]{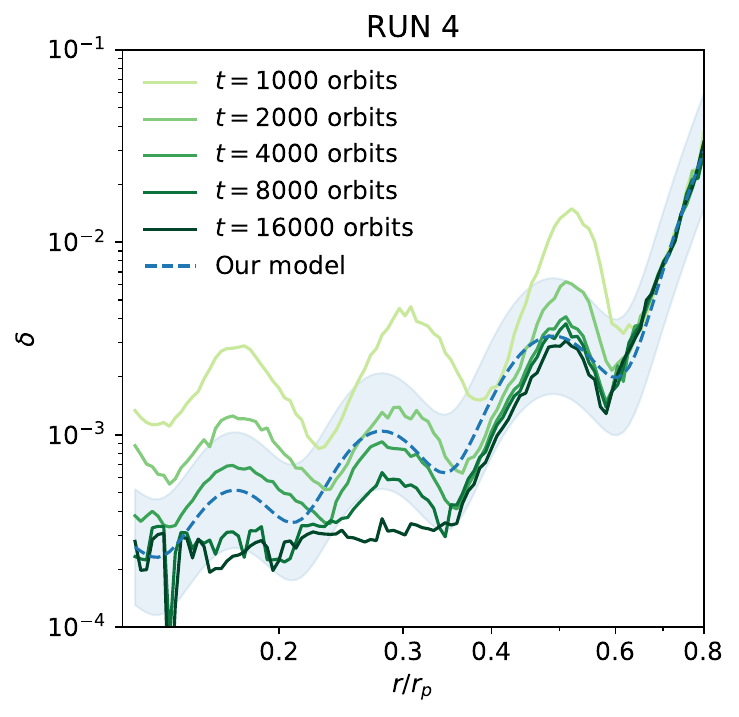} 
    \end{center}
    \caption{{Time evolution of entropy jump from the long-term simulations of RUN 4. The green lines represent snapshots. The blue dashed line indicates the empirical formula (equation (\ref{eq:fitting_entjump})). The shaded band indicates the deviation by factor of 2 from the empirical formula. }}
    \label{fig:delta_a35_Tcompare}
\end{figure}

{Figure \ref{fig:delta_a35_Tcompare} shows time evolution of the entropy jump profile from RUN 4. Although $\delta$ for the primary and secondary shocks has converged by 4000 orbits, $\delta$ for the tertiary shock decreases until 16000 orbits. We find that $\delta$ for the quaternary shock shows minor time evolution between 8000 and 16000 orbits, which is in contrast to the decreasing trend seen for the tertiary shock. 
We speculate that this is partly due to the detection limit of entropy jumps in our streamline analysis (see also sections \ref{sec:parameter} and \ref{subsec:simu_discus}). In such a case, the entropy jumps might be less accurate and overestimated. We defer further investigation of temporal changes of $\delta$ for such weak shocks to future work.
}

{Figure \ref{fig:delta_a35_Tcompare} implies that our empirical formula overpredicts entropy jumps for tertiary and quaternary shocks in the steady state in low-viscosity cases. Comparing the $\delta$ values measured at 16000 orbits to our empirical formula, we find that the difference is largest and a factor of three around the peak of $\delta$ for the tertiary shocks. This is slightly larger than the uncertainty of our empirical formula (a factor of two, see figure \ref{fig:delta_scaling}). The measured $\delta$ for the quaternary shock is within the error of our formula although the sinusoidal fitting might not be accurate. The deviation from the formula lies within the error range at 4000-8000 orbits, which approximately corresponds to 0.05--0.1 Myr at 5 au.
}

{
We thus suggest that our empirical formula with the uncertainty may overestimate entropy jumps achieved in the steady state for $\alpha\lesssim10^{-3.5}$. In such low-viscosity cases, one would need to take into account wind-driven angular momentum transport to model actual disk evolution \citep[e.g., see][for reviews]{Lesur23,Bai26}. Shock heating in wind-driven accretion disks will be addressed in our forthcoming paper (Tominaga et al. in prep.). 
}

\end{document}